\documentclass[%
reprint,
amsmath,amssymb,
aps,
]{revtex4-2}

\usepackage{graphicx}
\usepackage{dcolumn}
\usepackage{bm}
\usepackage{xcolor}

\begin{document}
	
	\preprint{APS/123-QED}
	
\title{Excitation Populations Provide a Thermodynamic Order Parameter for Liquids}
 
\author{Marcus T. Cicerone, Jessica Zahn, John P. Stoppelman, Jesse G. McDaniel}
\affiliation{Department of Chemistry and Biochemistry, Georgia Institute of Technology \\ 950 Atlantic Drive, Atlanta, GA 30332, USA}
\email{cicerone@gatech.edu}
\author{Kelly Badilla-Nunez}
\affiliation{School of Chemical and Biochemical Engineering, Georgia Institute of Technology \\ 900 Atlantic Drive, Atlanta, GA 30332, USA}

\date{\today}
	
\begin{abstract}

Simulation and model system studies suggest that local structural excitations play an important role in the dynamics of liquids and glasses. Here, for the first time, we quantify excitation populations in real liquids, showing that their temperature-dependent population can be predicted from entropy and enthalpy of melting. We further show that the excitation population in the first solvent shell serves as an order parameter for the appearance of dynamic heterogeneity and for driving transformations between distinct mechanistic regimes of liquid relaxation. We propose a scenario that provides simple physical explanations for these previously enigmatic aspects of liquid behavior. 
\end{abstract}

\maketitle
	
Relaxation phenomena in liquids reveal the presence of dynamic heterogeneity, temperature-dependent mechanistic crossovers, and, ultimately, ergodicity breaking at the glass transition. Perhaps the most striking of these is the strongly super-Arrhenius behavior of primary ($\alpha$) relaxation or viscosity. Goldstein \cite{Goldstein:1969eg} first proposed that neighbor-induced dynamic constraints in liquids could produce this behavior below some onset temperature. 


Nowadays, the onset temperature ($T_A$) is assigned to the point where the primary relaxation time ($\tau_\alpha$) crosses over from Arrhenius behavior, $\tau_\alpha \propto exp[E_A/k_B T]$, $E_A$ = constant for $T>T_A$, to super-Arrhenius behavior where $E_A \propto T^{-2}$. Below $T_A$, liquids evince non-ergodicity on the timescale of $\tau_\alpha$ through the time dependence of the rotational relaxation function $c(t)_R \propto exp(-t/\tau)^{\beta_{KWW}}$. Near $T_A$, $\beta_{KWW}$ generally begins to decrease from its high-temperature asymptotic value, signifying a broadening in the distribution of dynamic environments that persist on the timescale of $\tau_\alpha$.

Evidence for another crossover is found at a lower temperature  \cite{Das.2004,Mallamace:2010tg,Stickel:1995du}, that we denote as $T_B$. Here, $E_A$ takes on a stronger quadratic temperature dependence \cite{Stickel:1995du} and relationships between transport coefficients such as $\tau_\alpha$ and the diffusion coefficient ($D_T$) change abruptly \cite{Cicerone:1996tj, Qi.2000}. Additionally, $\beta_{KWW}$ values approach their asymptotic minimum, and there is evidence for the emergence of a dynamically inactive, solid-like phase \cite{Das.2004, Chandler:2010jm,Pinchaipat.2016, Das.2004}. Finally, at $T_B$, an additional relaxation ($\beta_{JG}$) \cite{Johari:1970dj} appears to bifurcate from the $\alpha$ process.
	
A potential energy landscape (PEL) formalism developed by Stillinger \cite{Stillinger:1995uj} as an elaboration of Goldstein's \cite{Goldstein:1969eg} proposal provides a framework for considering these dynamic phenomena. In this framework, dense, amorphous packing leads to locally favored particle configurations referred to as inherent states (IS). These lie in energy basins with width, depth, and complexity that are determined by the properties of the liquid particles. Thermodynamic properties are encoded in the PEL through statistics of basin distributions, and dynamic properties arise through interbasin (IB) transitions. It has been proposed that, below $T_A$, IB transitions are associated with excitations \cite{Chandler:2010jm}, localized structural defects that facilitate particle rearrangements.

Significant evidence exists for the importance of excitations. For example, the quadratic temperature dependence in $E_A$ just below $T_A$ is accounted for by excitation-facilitated release of dynamic constraints if relaxation events are predicated on the spatio-temporal intersection of two excitations \cite{Chandler:2010jm,Elmatad:2009ec}. At lower temperatures, near $T_B$, localized molecular hops associated with excitations have been identified with the $\beta_{JG}$ process \cite{Cicerone:2017eo, Yu:2017fw}, or, equivalently, the excess relaxation wing \cite{Guiselin.2021}. Excitation-mediated hops are thought to induce the dynamic crossover near $T_B$ \cite{Angell.2000kwg,Iwashita:PRL:2013}, preventing a divergence in $\tau_\alpha$ \cite{Das.2004}. Excitations have also been linked directly to density scaling \cite{Stoppelman.0} and plastic flow \cite{Cao.2019}. They are possibly linked to IS annealing behavior \cite{Nishikawa.2022} and unjamming \cite{Kapteijns.2018}. 
	
Owing to their putative centrality, excitations have been heavily investigated. They arise from packing frustration in amorphous systems \cite{Mizuno.2017}, and their population exhibits an $\omega^4$ frequency dependence \cite{Lerner.2016,Mizuno.2017} up to some cutoff $\omega_c$, above which it drops quickly \cite{Kapteijns.2018}. Their spatial extent is on the order of a few particle diameters \cite{Kapteijns.2018, Bhattacharyya.2002} and the hops they facilitate are spatiotemporally correlated, with loci forming spatially extended string-like structures at high temperatures \cite{Donati.1998} and compact structures below $T_B$ \cite{Stevenson.2006}. In simulations, their concentration (noted below as $\Phi_0$) follows a Boltzman distribution over the temperature ranges investigated \cite{Keys:2011er,Ortlieb.2021}. Because excitations represent saddle points between basins, $\Phi_0$ should be connected to the thermodynamic properties of the liquid. However, the nature of this connection has not been explored to our knowledge.

\begin{figure}
\includegraphics[width=0.45\textwidth]{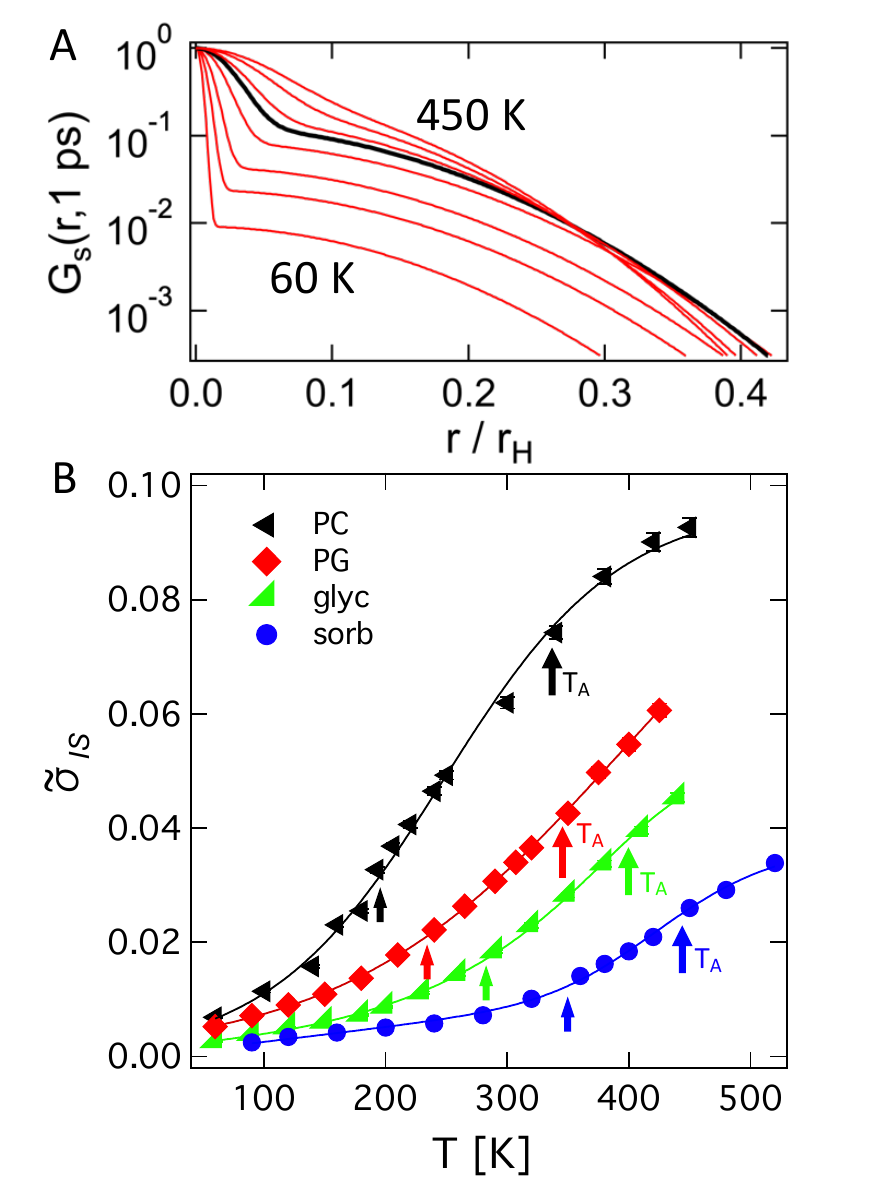}
\caption{\footnotesize{(A) Single particle van Hove function for propylene glycol from QENS at temperatures ranging from 60 K to 450 K at 55 K intervals. The solid black line marks $G_s$ at 290 K, 5 K above $T_A$. (B)  $\tilde{\sigma}_{IS}$ at 1 ps as a function of temperature for the liquids indicated. PC = propylene carbonate, PG = propylene glycol, glyc = glycerol, sorb = sorbitol. The 1 ps $\tilde{\sigma}_{IS}$ data are fitted by a generalized logistic function with a linearly increasing baseline. Fits to the data should be viewed as a guide to the eye.}\label{fig:vanSig}} 
\end{figure}

Excitations are typically identified in simulation and model systems through associated hops - distinctively large, rapid particle excursions \cite{Keys:2011er,Schoenholz:2016tz,Ortlieb.2021}. The spatial and temporal resolution required to identify these motions on a population basis is afforded in real liquids by incoherent quasielastic neutron scattering (QENS) \cite{Cicerone:2014dk,Cicerone:2017eo}. On the timescale of hopping events, roughly 1 ps, \cite{Cicerone:2017eo,Schoenholz:2016tz,Ortlieb.2021,Bender:2020dl}, diffusion is not significant. Thus, as discussed in Supplementary Material \footnote{See the Supplemental Material at URL for details on fitting the QENS data, Simulation details for finding first-shell coordination numbers, Tables of $T_A$ and $T_B$ values, and details related to simulation and analysis of the Kob Andersen (KA) Lennard-Jones model system.} we can approximate the self-intermediate scattering function from QENS as \cite{Cicerone:2017eo}:
\begin{equation}
	F_s(q,t)=(1-\Phi(t))e^{-(q\,\pi\,\sigma_{IS}(t))^2}+\Phi(t) e^{-(q\,\pi\,\sigma_{IB}(t))^2} 
	\label{eq:F2Gauss}
\end{equation}	
where $\Phi(t)$ gives the fraction of particles involved in an excitation-mediated IB barrier crossing up to time t. $\sigma_{IS}$ is the characteristic lengthscale for elastic deformations of particles about their fiduciary or inherent state positions. $\sigma_{IB}$ is the characteristic lengthscale for an interbasin transition; the particle motion over a saddle point leading to conversion from one IS to another \cite{Cicerone:2017eo}. Fits of Eq. (\ref{eq:F2Gauss}) to $F_s(q,t)$ data are shown in Supplemental Material. Figure \ref{fig:vanSig}A shows the space Fourier transform of $F_s(q,t)$, the single particle van Hove function $G_s(r,t=1ps)$ for propylene glycol (PG) over a temperature range from (60 to 450) K. The distinguishability between IS and IB motion is evident from the bi-modal nature of $G_s$. The lengthscale of the more localized mode is shown to be strongly temperature dependent in Fig. \ref{fig:vanSig}B where we plot $\tilde{\sigma}_{IS}=\sigma_{IS} / r_H$ ($r_H$ is the hydrodynamic radius) at 1 ps for the four liquids indicated. The fact that the amplitude of the less localized mode in $G_s$ changes strongly with temperature, but its shape does not shows that $\sigma_{IB}$ is more-or-less temperature-independent but that its population $\phi$ depends strongly on temperature. (More on this below.)
	
The heavy black line in Figure \ref{fig:vanSig}A gives $G_s(r,t)$ at $T_A$. The fact that IB barrier crossing motion remains statistically distinct from IS motion at 1 ps, even well above $T_A$ suggests that relaxation is neither truly collisional nor unaffected by neighbor-induced dynamic constraints as typically assumed \cite{Chandler:2010jm,Stevenson.2006,Bhattacharyya:2008ux}. Rather, even at $T\geq T_A$, a sizable fraction of molecules remain trapped in their respective basins at short times. This interpretation of the data in Fig. \ref{fig:vanSig}A idea is reinforced by the $\tilde{\sigma}_{IS}$ values in Fig. \ref{fig:vanSig}B. Assuming basins with quadratic potentials, $>85$\% of particles will remain caged when $\tilde{\sigma}_{IS}$ is $<80$\% of its asymptotic maximum, as they appear to be at $T_A$. (See Supplemental Material).


\begin{figure*}
\includegraphics[width=1\textwidth]{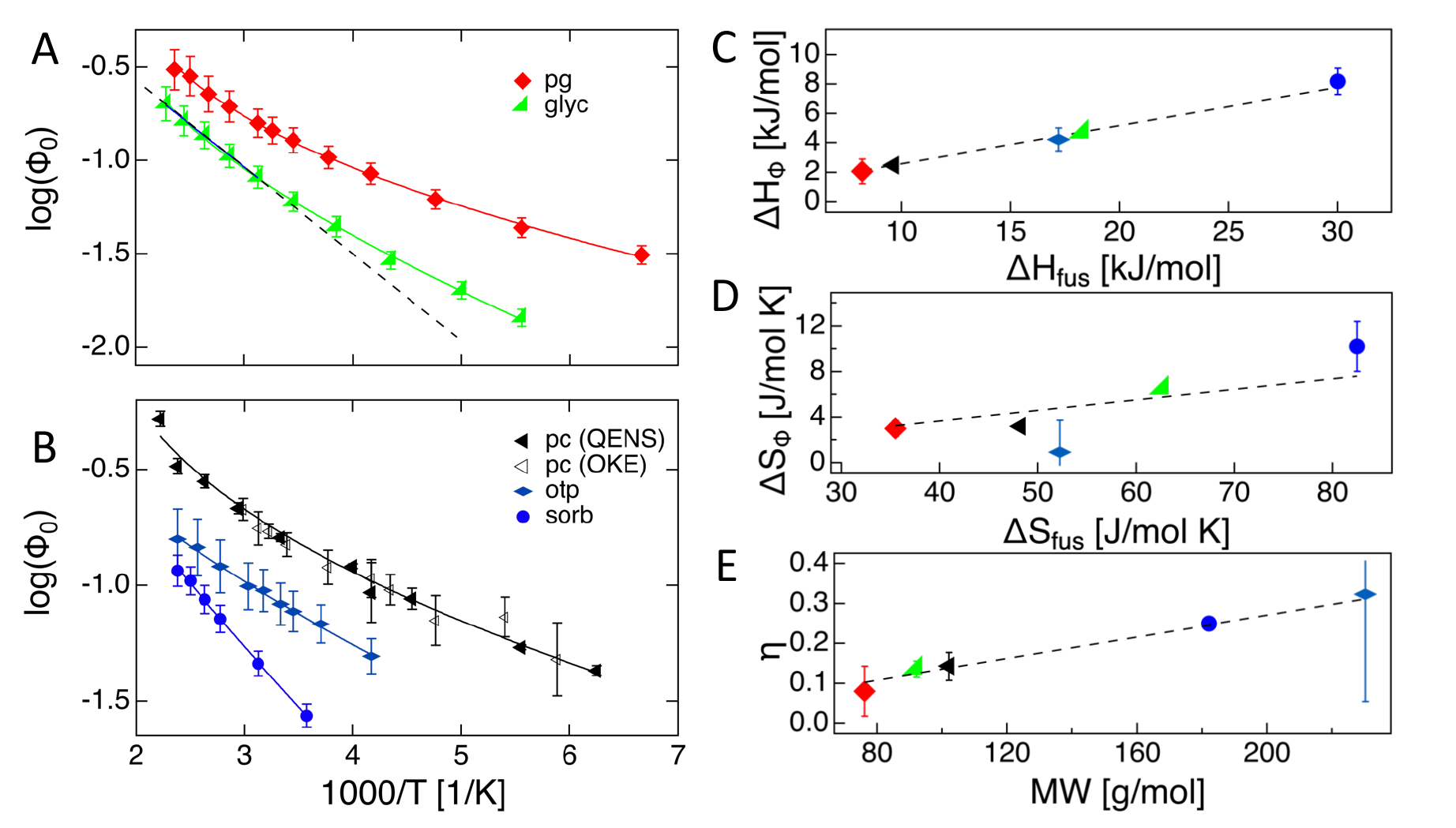}
\caption{\footnotesize{Panels A \& B: Instantaneous excitation populations, $\Phi_0$ as a function of inverse temperature for the liquids indicated. PC was measured also by optical Kerr effect \cite{Bender:2020dl}. Solid lines are fits to Equation 1. The data are presented in two panels for clarity. The dashed black line in panel A is a Boltzmann fit to the glyc data at $T>T_B$. Panel C, D \& E: Fit parameters for data in panels A \& B to Eq (1), enthalpy and entropy change for creation of excitations are plotted against $\Delta H$ and $\Delta S$ for crystallization. Symbol-substance associations in panels C, D \& E are the same as in panels A \& B. Dashed lines are best fits to the data with a constraint of intercept at the origin.}\label{fig:Phi0}}
\end{figure*}
 
Below, we consider the excitations that ostensibly facilitate the release of these constraints. We will initially be interested in the instantaneous population of excitations ($\Phi_0$), which gives the thermodynamically controlled population at saddle points. We obtain $\Phi_0$ from $\Phi(t\approx 1ps)$ when particles have just explored their constraints \cite{Larini:2008ur} and $\tilde{\sigma}_{IB}$ reaches a local maximum (see Supplemental Material). This timescale and the lengthscale of $\tilde{\sigma}_{IB}(1ps)\approx 0.2\, r_H$ are consistent with criteria applied in simulation and colloid experiments for identifying excitations \cite{Keys:2011er,Schoenholz:2016tz,Ortlieb.2021}.

Panels A and B of Figure \ref{fig:Phi0} show values of $log(\Phi_0)$ as a function of inverse temperature for the four molecular liquids from Fig. \ref{fig:vanSig}B and ortho-terphenyl. Published excitation population data that we know of are derived from classical dynamics simulation \cite{Chandler:2010jm, Keys:2011er, Ortlieb.2021}, and are consistent with a Boltzmann distribution over the temperature range they cover. The dashed black line in Panel A is a Boltzmann law ($\Phi_0\propto Exp[-\epsilon/k_B T]$) fit to the Glycerol $\Phi_0$ at $T\geq T_B$, with $\epsilon=9.1 kJ/mol$. This fit is reasonable at high temperatures, but it underestimates $\Phi_0$ below $\approx 300\,K$. We cannot determine whether this deviation is due to temperature dependence in the partition function or the onset of quantum effects associated with bosonic excitations. We leave this question to future work and note that the data are consistent with Bose-Einstein (BE) statistics as indicated by the solid-line fits to:
\begin{equation}
    \Phi_0=\frac{\eta}{exp\left[\Delta G_\Phi/k_B T\right]-1} \label{eq:BE}
\end{equation}
where $\Delta G_\Phi = \Delta H_\Phi-T\Delta S_\Phi$ is the free energy cost, and $\Delta H_\Phi$ and $\Delta S_\Phi$ are the enthalpy cost and entropy gain in creating an excitation. Eq. (\ref{eq:BE}) describes the population of a single state, while a distribution of states, $\Phi_0(\omega)\propto \omega^4$ is expected \cite{Lerner.2016,Mizuno.2017}. However, this distribution is strongly peaked at its cutoff, $\omega_c$ \cite{Kapteijns.2018}, so we may assume that our data primarily reflects the temperature dependence of the distribution at $\omega_c$. 

The dashed lines in panels C through E of Figure \ref{fig:Phi0} are best linear fits with intercepts at the origin plotted against enthalpy and entropy of fusion, respectively. Thermodynamic parameters for excitation formation plotted in panels C and D are obtained from fits to Eq. (\ref{eq:BE}). Regarding their correspondence to melting parameters, it is notable that similar excursion lengths are used to distinguish excitations \cite{Ortlieb.2021} and local melting events \cite{Chakravarty.2007} in Lennard-Jones models. We find that $\Delta H_\Phi \approx 0.3\,\Delta H_{fus}$, whereas $\Delta S_\Phi \approx 0.1\,\Delta S_{fus}$. The relatively smaller proportionality for entropy likely reflects a more localized and slightly less collective nature of excitation formation than that of melting. The proportionality of $\eta$ to molecular weight (MW) in Fig. \ref{fig:Phi0}E is indistinguishable from a linear dependence on the number of H atoms/molecule and likely reflects a (yet unexplored) role of intramolecular bonds on the spatial extent of excitations. Given the range of intermolecular interactions that dominate in these liquids, we expect the relationships exemplified in Figure \ref{fig:Phi0} to hold for molecular and atomic liquids in general, making it possible to estimate excitation populations simply from melting thermodynamics. 
	
\begin{figure*}
\includegraphics[width=1\textwidth]{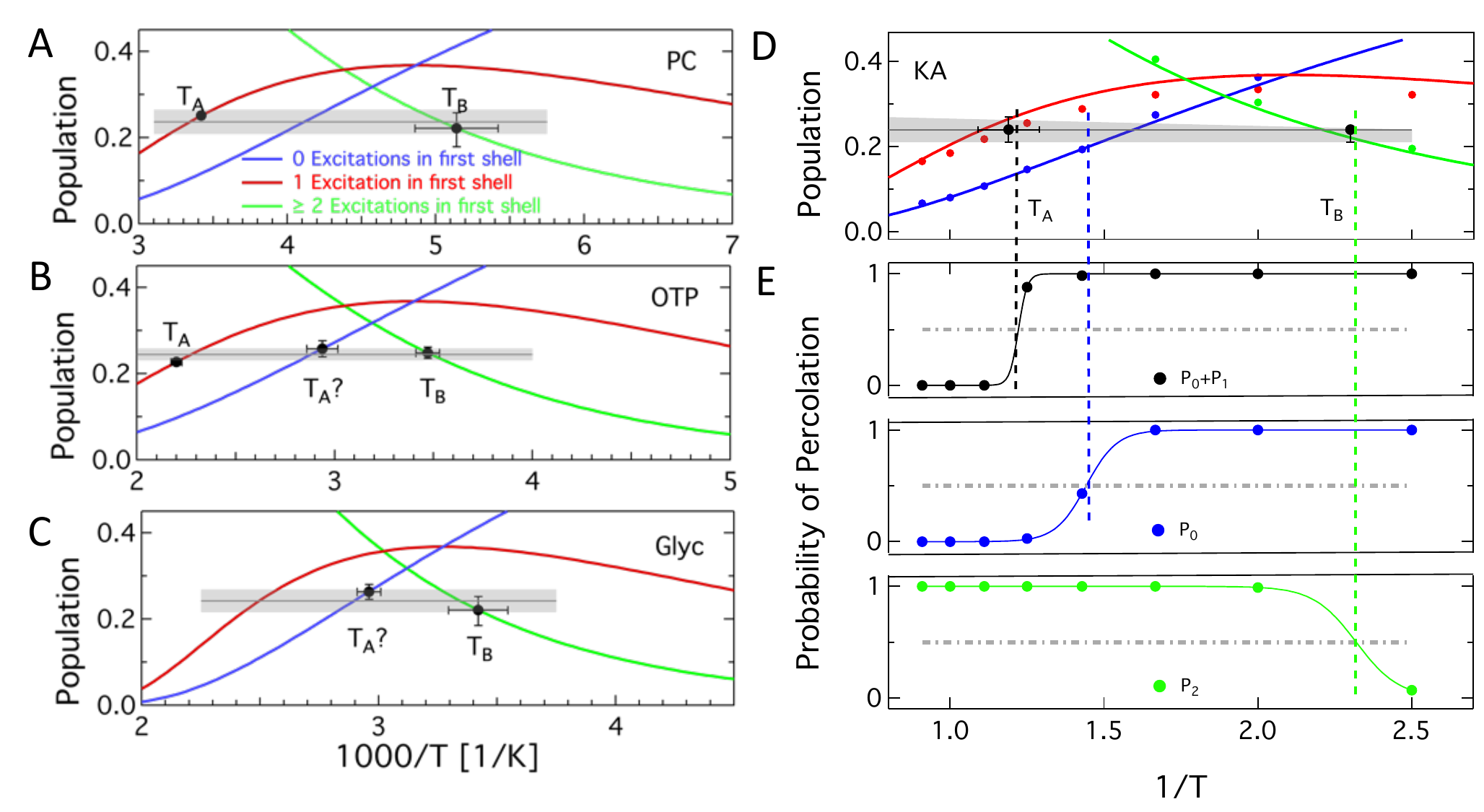}
\caption{\footnotesize{(A-D) Estimated populations of environments with 0, 1, and $\geq 2$ excitations in the first shell as a function of temperature for liquids indicated and the KA model. Black circles mark $T_A$ and $T_B$ values. $T_A$ and $T_B$ always coincide with a population of $0.24\pm 0.02$ for $P_2$, $P_1$, or $P_0$. (D) Blue, red, and green circles mark $P_0$, $P_1$, and $P_2$ populations calculated directly from trajectories. (E) Black, blue, and green circles give the probability that ($P_0$+$P_1$), $P_0$ alone, and $P_2$ domains that persist for $\approx$2.5 ps in KA will form percolating clusters.}\label{fig:regimes}}
\end{figure*}
			
Having $\Phi_0$ data for these liquids from well above $T_A$ to well below $T_B$, we are in a position to explore how excitation-induced constraint release might impact relaxation mechanisms. Assuming that dynamics are influenced primarily by the structure of the first solvation shell \cite{Schoenholz:2016tz,Tong.2019}, we classify local dynamic environments by the number of molecules within the first shell involved in excitations. Retaining the assumption of facilitated dynamics, we argue that only three distinct environments are relevant; those having zero, one, or $\geq$2 excitations in the first coordination shell. We indicate populations in these environments as $P_0$, $P_1$, and $P_2$.

	
We estimate the populations of these distinct environments by assuming that excitations are randomly distributed in space and that $P_0= 1$ at $T=0$. Under these assumptions, $dP_0/d\Phi_0=-(z+1)\,P_0$, where z is the coordination number, and $dP_i/d\Phi_0=(z+1)\,(P_{i-1}-P_i)$ for $i>0$. In Figure \ref{fig:regimes} and the Supplementary Material, we plot estimated values of $P_0$, $P_1$, and $P_2$ for all liquids analyzed here and for a KA model. The circles in Fig. \ref{fig:regimes}D show environment populations calculated directly from simulation trajectories. (See Supplementary Material for simulation details). Although excitations are inhomogeneously distributed in the KA model \cite{Keys:2011er}, our $P_i$ population estimations are quite close to measured values in the KA system. Aside from the assumption of random distribution, the populations are estimated with no free parameters. $\Phi_0$ is determined experimentally, and z is calculated from simulation (see Supplemental Material).

$T_A$ and $T_B$ values shown in Fig. \ref{fig:regimes} and their uncertainties are tabulated from literature. (See Supplemental Material.) At each of these temperatures, one of the populations $P_i$ has a value $0.24\pm0.02$. Vertical error bars at these intersections are obtained by projecting temperature uncertainty onto the associated population line. In most cases, $P_1$ and $P_2$ = $0.24$ within uncertainty at $T_A$ or $T_B$ respectively. For OTP, glycerol, and PG, transport anomalies have also been reported at temperatures where $P_0=0.24$, and have been assigned, perhaps erroneously, to $T_A$ or $T_B$.

The probability that a single characteristic temperature for any individual liquid would randomly fall at $P_i=0.24$ is approximately equal to the uncertainty in that $P_i$ value ($2\sigma_i$) at that temperature, divided by the total range of $P_i$. Accordingly, we estimate the log probability of the null hypothesis for N such temperatures as $ln(H_0)\approx N\,ln[2 \sigma_i/range(P_i)]$ where we generously assign $range(P_i)=0.35$. $H_0$ values for individual liquids are modest, but $H_0\approx10^{-8}$ for all systems combined. 

The data of Figure \ref{fig:regimes}A-D appear to suggest that percolation of dynamically distinct environments underlies dynamic crossovers. However, $0.24\pm0.02$ is too high for a site percolation threshold ($p_c$) in these systems. These random media should exhibit $p_c$ values below $p_c=0.2$ for a regular lattice \cite{Ziff.2017} with the same z=12. Below, we resolve this conundrum through a deeper analysis of the KA system. We have previously demonstrated that on timescales and lengthscales associated with interbasin hops, motion in molecular systems can be mapped onto behavior of simple systems such as the KA model \cite{Cicerone:2017eo,Stoppelman.0}. 

Fig. \ref{fig:regimes}E shows that $T_A$ and $T_B$ in the KA model mark percolation thresholds at $p_c=0.14$ for environments that persist for at least 2.3 ps (see also Supplemental Information). So far, we have defined excitations and environments $P_i$ at $\approx$ 1 ps when hops are first definable, but, consistent with our analysis below, we postulate that environments which are too short-lived will not have a significant impact on dynamics.

We can qualitatively understand the major dynamic features of liquids as direct consequences of the prevalence and persistence of excitations in local environments. Assuming that excitations are required for relaxation, $P_0$ environments are solid-like and unable to relax, and $P_1$ environments support only single IB transitions (e.g., two-level systems). In $P_2$ environments, reversible IB transitions can be entropically trapped by a quasi-synchronous and nearby IB transition, leading to non-reversing relaxation events. Thus, while the excitations themselves are bosonic and do not interact directly \cite{Nishikawa.2022}, \textit{they interact indirectly through their effects on the material}.

In extended $P_2$ environments, elemental relaxation events may be as short as the interbasin barrier crossing time, $\approx 1\,ps$. Nearby relaxations would rearrange the local structure over the hop duration, accounting for the reduced hop length at elevated temperatures \cite{Cicerone:2014dk}, and defining small-step diffusion as instantaneously facilitated relaxation dynamics that will be observed only in extended $P_2$ domains, i.e., only for $T>T_B$ \cite{Bhattacharyya.2002,Das.2004}. 

In the proposed scenario, both $P_0$ and $P_1$ exist at $T>T_A$, but neither forms extended domains. The existence of slow environments that persist on a ps timescale is demonstrated by the bimodal structure in Figure \ref{fig:vanSig}A. The fact that these environments are transient on the timescale of $\alpha$ relaxation is evinced by the Arrhenius temperature dependence and lack of obvious dynamic heterogeneity signatures. We understand this to result from $P_0$ and $P_1$ existing primarily as individual loci or small droplets surrounded by $P_2$ at $T>T_A$. Intimate proximity with $P_2$ environments means that $P_1$ or $P_0$ particles will quickly become involved in an excitation, leading to the interconversions $P_0 \leftrightarrow P_1 \leftrightarrow P_2$ on the timescale $\tau_\Phi$, where $\tau_\Phi < 0.25 \left<\tau_\alpha\right>$ for $T\geq T_A$ \cite{Cicerone:2017eo}. Interconversion of these environments on such a short timescale would obfuscate dynamic heterogeneity effects, and $\alpha$ relaxation would \textit{appear} to be homogeneous and collisional.

At lower temperatures, remaining halmarks of liquid dynamics can also be understood in the context of our model. At $T<T_A$, signatures of dynamic heterogeneity and excitation-facilitated hops become detectable \cite{Das.2004,Chandler:2010jm,Mirigian.2014}, which we understand as a consequence of increased persistence of slow domains ($P_0$ plus $P_1$) concomitant with their percolation. The appearance of extended $P_0$ plus $P_1$ environments is consistent with expectations of a spinodal at this temperature \cite{Stevenson.2006}. Below $T_B$, small-step diffusion disappears \cite{Bhattacharyya.2002,Das.2004}, which we understand as the loss of extended $P_2$ domains at this same temperature. The existence of extended $P_2$ domains only at $T>T_B$ is also consistent with the observation that saddle points are encountered when quenching simulations to identify inherent states from this temperature regime \cite{Nishikawa.2022}, and with a localization transition (loss of saddle points) at $T<T_B$. The loss of extended $P_2$ domains would eliminate the component of non-reversing relaxation that can occur on the timescale of IB barrier crossing, leading to the bifurcation of the former from the latter - i.e., the bifurcation of the $\alpha$ and $\beta_{JG}$ process.

We also note that, below the temperatures where $P_0$ percolates, solid-like domains should begin to persist long enough to impact dynamics. Accordingly, there is a shift in the temperature dependence of the Brillouin line at precisely this temperature in PC \cite{Brodin.2002}. Additionally, this temperature seems to have been mistakenly identified as $T_A$ or $T_B$ in OTP, glycerol, and PG. Finally, two-level systems are known to disappear in glasses prepared with a sufficiently low equilibrium structure \cite{Khomenko.2020}. This is predicted in the scenario proposed here when $P_1$ drops below the detection limit at some temperature well below $T_B$. 


\begin{acknowledgments}
We thank Jack Douglas, David Simmons, and Mark Ediger for their insightful comments and suggestions. 
We acknowledge the support of the National Institute of Standards and Technology, U.S. Department of Commerce, in providing the neutron research facilities supported in part by the National Science Foundation under Agreement No. DMR-0454672. 
\end{acknowledgments}

%

\end{document}


\preprint{APS/123-QED}
	
\title{Excitation Populations Provide a Thermodynamic Order Parameter for Liquids}
 
\author{Marcus T. Cicerone, Jessi Zahn, John P. Stoppelman, Jesse G. McDaniel}
\affiliation{Department of Chemistry and Biochemistry, Georgia Institute of Technology \\ 950 Atlantic Drive, Atlanta, GA 30332, USA}
\email{cicerone@gatech.edu}
\author{Kelly Badilla-Nunez}
\affiliation{School of Chemical and Biochemical Engineering, Georgia Institute of Technology \\ 900 Atlantic Drive, Atlanta, GA 30332, USA}

\date{\today}

\maketitle

\section*{Methods}

\subsection*{QENS Data Acquisition and Reduction}

Acquisition of QENS data analyzed here has been reported previously \cite{Cicerone:2014dk, Cicerone:2017eo}. In summary, the QENS data was acquired with the Disk Chopper Spectrometer (DCS) installed at the NG4 guide at the NIST Center for Neutron Research \cite{Copley:2003es}. The neutron wavelength was set at 4  \AA, corresponding to a momentum transfer (q) range of 0.2 \AA$^{-1}$ to 2.8 \AA$^{-1}$, and data were binned at steps of 0.1  \AA$^{-1}$. The instrumental resolution was 0.19 meV and the maximum energy transfer was 4.5 meV. Data at each temperature was obtained within (3 to 6) h. 

Experiments were performed on cooling from the high temperature to avoid potential artifacts from crystal nucleation. We ensured that there was no sign of crystallization even on reheating when crystal nuclei would have grown if they had been seeded at low temperatures. The absence of crystallization was confirmed by noting an absence of Bragg peaks in plots of scattered intensity vs q.  

The data were corrected for i) background from an empty can ii) dark count background with no neutron flux and iii) detector efficiencies by measuring a vanadium standard. All the data files were reduced using the DAVE software available at \underline{www.ncnr.nist.gov/DAVE}. Instrument resolution was estimated as a Gaussian from sample scattering at 30 K.

QENS data were analyzed in the frequency domain and in the time domain after transformation. $S(q,\omega)$ was directly transformed to the time domain $F(q,t)$ using DAVE software. Noise is propagated in the Fourier transform operation, and additional errors can arise through truncation and course sampling intervals. These factors are taken into account and standard uncertainties are calculated by the DAVE software. Errors arising were less than 0.1\% of $F(q,t)$ values. The real part of the Fourier transformed data was used to calculate $F(q,t)$, and no filtering options were used as they were not found necessary. 

\subsection*{Modeling QENS Data}

In this work, we use an explicitly heterogeneous dynamics model to describe the QENS data. We account for distinct types of motion expected in amorphous systems: over-damped vibration (IS motion), hops (IB transitions), and homogeneous diffusion. Based on ultrafast optical experiments performed by us \cite{Bender:2020dl}, and on work by Vispa et al.,\cite{Vispa.2015} we use Lorenzians to represent each of the three types of motion considered here. Vispa et al.,\cite{Vispa.2015} found that a triple Lorentzian function provided significantly better fits for $S(q,E)$ of a molecular liquid than common models of similar complexity containing functional forms such as KWW and Gaussian, and much better fits than 2-component models. Likewise, we find that a three-Lorenzian model fits scattering data well for each of the liquids analyzed here. Consistent with this, we observe three exponential relaxation processes for PC in time-domain optical Kerr effect data covering similar time and lengthscales to those considered here. Accordingly, we use the following model for $S(q,E)$:
\begin{align}
	S(q,E)=&(1-\tilde{\Phi}(E))L_D\otimes[(1-a_v)\,\delta(E)+a_v L_v]\nonumber\\
	&+\tilde{\Phi}(E) L_D\otimes[(1-a_v)\,\delta(E)+a_v L_v]
	\otimes[(1-a_h)\,\delta(E)+a_h L_{h}]\label{eq:FullSQE}
\end{align}
where $L_i=\Gamma_i\pi^{-1}(E^2+\Gamma_i^2)^{-1}$, $\otimes$ is the convolution operator, and the convolutions are over frequency (energy). $\Gamma_D=D_Tq^2$ where $D_T$ is the diffusion coefficient, and $a_v$ and $a_h$ are respectively, q-dependent scattering amplitudes from vibration (IS motion) and hopping (IB crossing). Anticipating that these two will be localized modes with Gaussian distributions of displacements,\cite{Rahman:1962ed} we assume the functional form:
\begin{equation}
	a_i(q)=c_i\{1-exp[-(\pi\sigma_i q)^2]\}\label{eq:LocMod}
\end{equation}
where the $\sigma_i$ represents the characteristic lengthscales of motion for mode $i$.

The two terms in Eq (2) account for two dynamically different classes of molecules. We assume that all molecules undergo both diffusion and over-damped vibrations at all times. Accordingly, both terms in Eq. (2) include diffusive and vibrational motion. We also assume that, up to a given time, only some fraction ($\Phi$) of molecules have participated in collective hopping motion ($h$) associated with IB transitions. These are accounted for in the second term of Eq. (2).

Data are first fitted in the frequency domain to obtain q-dependent characteristic frequencies $\Gamma$ for each mode of motion. Data are then transformed to the time domain $F(q,t)$ and fitted to the time-domain representation of Eq. (2) to obtain $\Phi (t)$:

\begin{equation}
	F(q,t)=e^{-t/\tau_D}[1+a_v(e^{-t/\tau_v}-1)]\\ 
	\times [(1-\Phi(t))+\Phi(t)(1+a_h(e^{-t/\tau_h}-1))] \label{eq:F2GFull}
\end{equation}
where $\tau_i=\hbar/\Gamma_i$. In the regime $t\Gamma_D \ll\hbar \ll t\Gamma_h\leq t\Gamma_v$, we can ignore terms involving $\tau_v$ and $\tau_h$, and Eq. (\ref{eq:F2GFull}) reduces essentially to
\begin{equation}
F(q)=(1-\Phi(t))e^{-(q\,\pi\,\sigma_{v})^2}+\Phi(t) e^{-(q\,\pi\,\sigma_{h})^2} 
\label{eqn:F2GShort}
\end{equation}

We have used to Eq. (\ref{eq:F2GFull}) fit QENS when long-time response is of interest \cite{Cicerone:2017eo}, and  Eq. (\ref{eq:F2GFull}) when only short time ($\approx 1\,ps$) is of interest \cite{Cicerone:2014dk}. In either case, $\Phi(t)$ represents the fraction of molecules that have hopped (undergone an IB crossing) up to time t. Bearing in mind that hops constitute significantly larger excursions than vibrations, it is clear that until a molecule executes a hop, the q-dependence of its scattering signature will be characteristic of only small lengthscale motion. Once a hop occurs, that signature will change to a larger lengthscale for all subsequent times. 

The QENS data, as obtained in the frequency domain, is a convolution of the molecular scattering function and instrument resolution
\begin{equation}
	S_{expt}(q,E)=S_{molecular}(q,E)\otimes R(q,E)
\end{equation}
where $R(q,E)$ is instrumental resolution. The model was convolved with the instrument response function when fitting in the frequency domain. The frequency-domain convolution operation is simple multiplication in the time domain: 
\begin{equation}
	F_{expt}(q,t)=F_{molecular}(q,t)\times R(q,t)
\end{equation}
where $R(q,t)$ is the Fourier transform of $R(Q,E)$. Thus, the molecular response was directly deconvolved from the overall signal as $F_{molecular}= F_{expt}/R$.

\begin{figure}[h!]
	\begin{center}
		\includegraphics[width=9 cm]{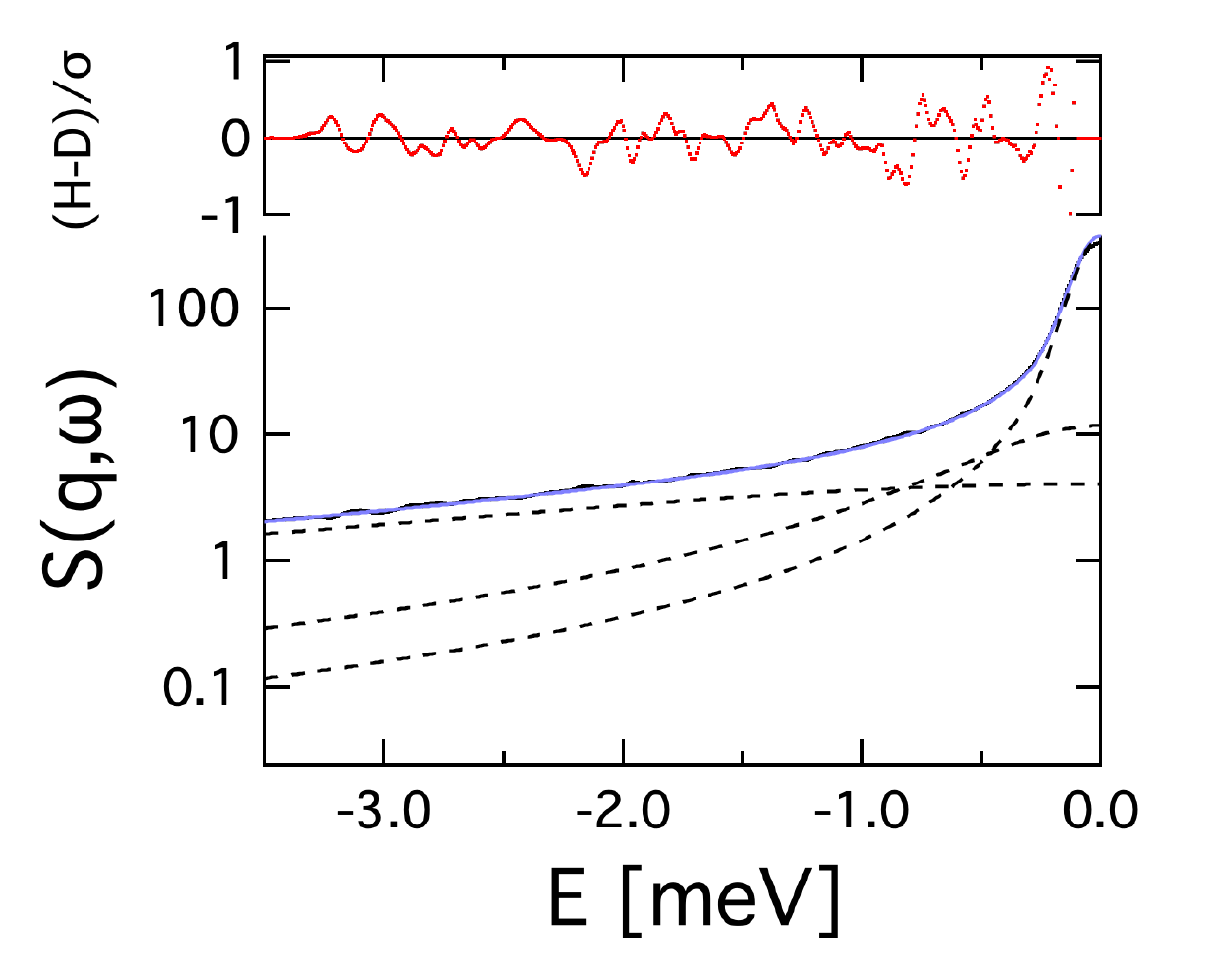}
		\caption{Fits to $S(q=0.8$\AA$^{-1},E)$ of PC, measured at T=300 K, using Eq. (2). The solid black line is the QENS data, $S(q,E)$. The solid blue line is the fit. The dashed lines are the fit components from three Lorenzians with $\Gamma_D=0.05$ meV,  $\Gamma_{IB}=0.55$ meV, and $\Gamma_{IS}=2.9$ meV. Fit residuals, normalized to uncertainty in the data points, are shown in the upper part of the figure.}\label{fig:SQWwFit}
	\end{center}
\end{figure}

$S(q,E)$ obtained at 4 K (or its Fourier transform) was used as the resolution function $R(q,E)$. Scattering from samples at 4 K was compared with standard vanadium and found to contain no broadening from, e.g. methyl group rotation, etc., so these data sets were deemed acceptable for normalizing $F(q,t)$. The fact that coherent scattering is only about 5\% of the total scattering in these samples ameliorates the complications arising from coherent scattering when base temperature data is used as a resolution for higher temperatures.

\begin{figure}[h!]
	\begin{center}
		\includegraphics[width=13 cm]{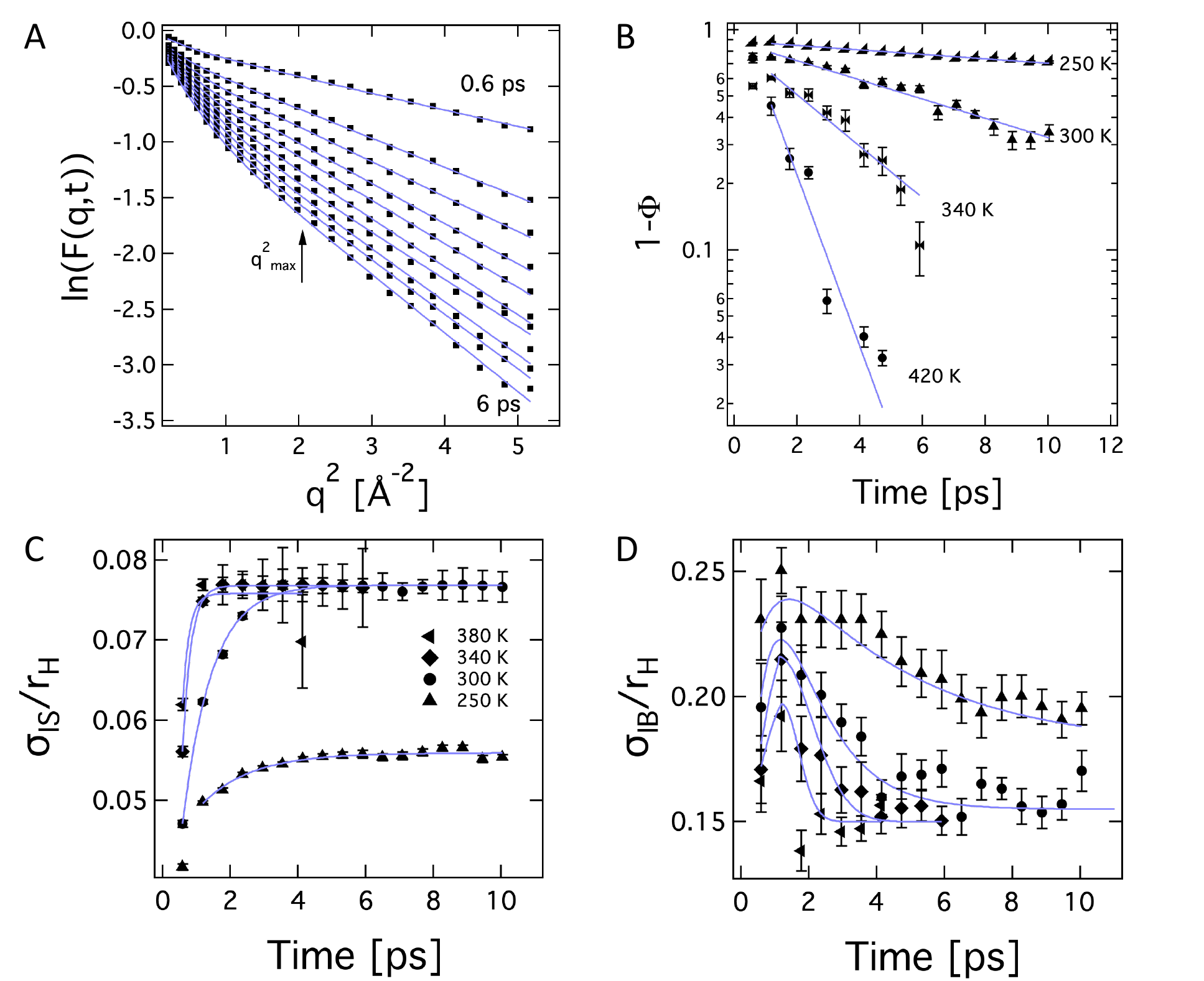}
		\caption{$F(q,t)$ from PC at 300 K and fits to Eq. (4). (A)Fits to $F(q,t)$ calculated at times ranging from 0.6 ps to 6 ps in increments of 0.6 ps. Solid lines are fits to Eq. (4). The uncertainties in the data are approximately the size of the symbols. (B) Time  dependence of  $\Phi$, derived  from  fits  of  Eq. (4) to F(q,t)  for temperatures indicated. Dashed lines are exponential decays. (C-D) Time dependence of  $\tilde{\sigma}_{IS}$ and $\tilde{\sigma}_{IB}$ derived from fits of Eqs. (3) and (4) to F(q,t) data in the range (250–380) K for PC. The solid lines are guides to the eye. Error bars indicate uncertainties in parameters at one standard deviation.}\label{fig:FqtFits}
	\end{center}
\end{figure}

In both frequency-domain and time-domain fitting, we used an iterative simulated annealing algorithm implemented in PyMC (\underline{pymc.io}), a Bayesian analysis framework, to find optimized fit parameters. In the frequency domain, we found values for $\Gamma_i$ at each average q value. Figure \ref{fig:SQWwFit} shows the fit of Eq. (2) to quasielastic neutron scattering (QENS) from PC at 300 K and q= 0.8 \AA$^{-1}$. The model is convolved with the instrument resolution. The upper trace shows fit residuals (H-D), normalized by the standard deviation (sd) at each data point. These normalized residuals are randomly distributed between -1 and 1 for all data reported, as in Fig. \ref{fig:SQWwFit}.

We used Eq. (\ref{eq:F2GFull}) to fit the time domain data shown in Fig.\ref{fig:FqtFits}, finding optimized parameters for $\Phi(t)$ and from Eq. (\ref{eq:FullSQE}), $c_i$ and $\sigma_i$. Figure \ref{fig:FqtFits}A shows fits of Eq. (\ref{eq:F2GFull}) to $F(q,t)$ from PC at 300 K at a series of times from 600 fs to 6 ps. Figure \ref{fig:FqtFits}(B-D) shows the time dependence of $\Phi$, $\sigma_{IS}$, and $\sigma_{IB}$ respectively at the temperatures indicated. 

$\Phi$, $\sigma_{IS}$, and $\sigma_{IB}$ each show time dependence consistent with their assignments. $\sigma_{IS}$ reaches an asymptote consistent with a localized process. $\sigma_{IB}$ behaves non-monotonically as expected if hop events have a non-zero probability of reversal. $\Phi$ increases monotonically from $\approx 0$ to 1.

\newpage

\subsection*{Fraction of Particles Caged}

\begin{figure}[h!]
	\begin{center}
		\includegraphics[width=15 cm]{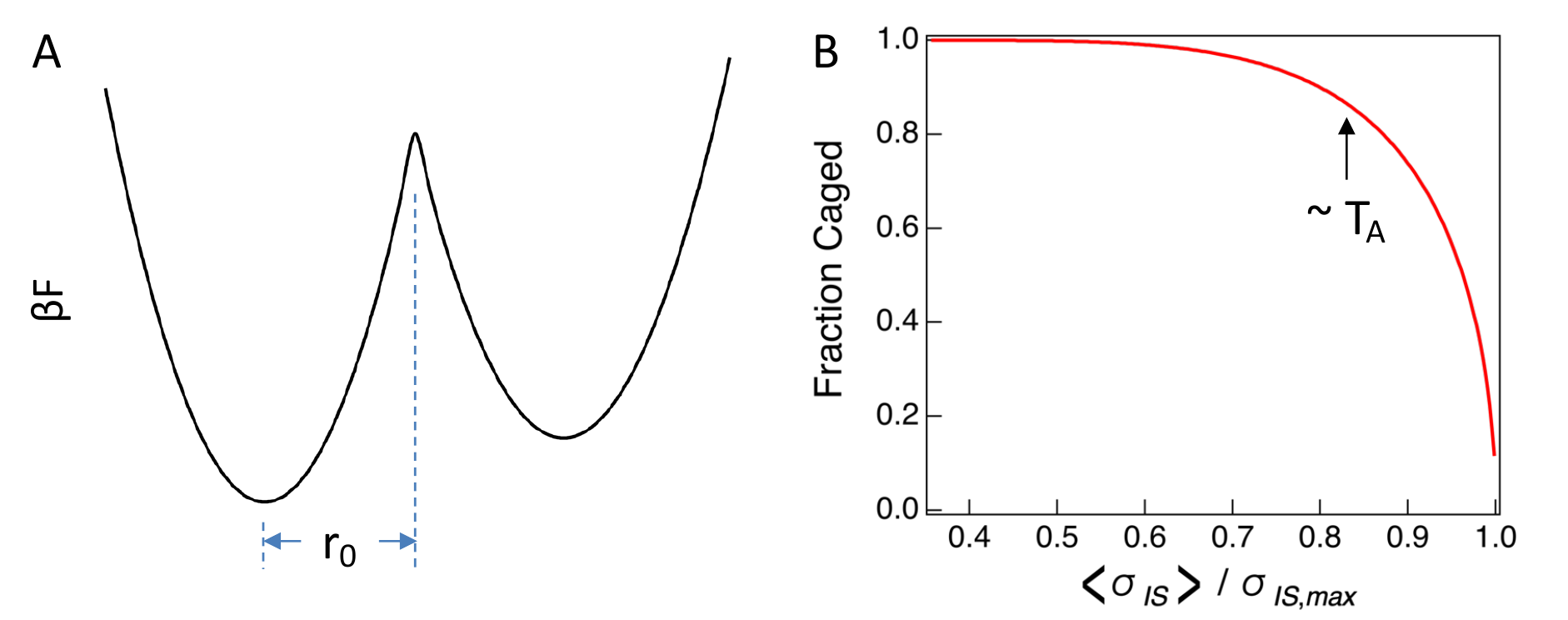}
		\caption{A: A free energy surface. B: Fraction of molecules instantaneously caged as a function of the characteristic lengthscale of IS elastic deformation $\sigma_{IS}$, normalized to $r_0$.}\label{fig:Caging}
	\end{center}
\end{figure}

We can estimate the fraction of particles caged as a function of the characteristic length for IS motion, $\sigma_{IS}$ by assuming a basin with free energy that is quadratic in a generalized displacement coordinate $r$ away from an inherent state energy minimum located at $r=0$, $F=r^2$, as sketched in Fig \ref{fig:Caging}A. Bearing in mind that our analysis of $F_s(q,t)$ categorizes only short lengthscale bound motion as IS deformations, $\sigma_{IS} \leq r_0$ where $r_0$ is the distance to the an interbasin barrier maximum, since larger excursions would lead to transit of the saddle point and be classified as IB motion. Under these assumptions, we can find the fraction of particles instantaneously caged, i.e., the population of particles undergoing excursions with $\sigma_{IS} < r_0$ as 
\begin{equation}
    f_{caged} = p(\sigma < r_0)=\frac{\int_0^{r_0} Exp[\beta F(r)] dr}{\int_0^\infty Exp[\beta F(r)] dr}=erf(r_0\sqrt{\beta} ) 
\end{equation}
where $\beta$ is an inverse energy scale.

The average excursion lengthscale for IS motion, $\left<\sigma_{IS}\right >$ is then given as the first moment of excursion lengthscales for caged particles:

\begin{equation}
    \left<\sigma_{IS}\right >=\frac{\int_0^{r_0} r Exp[\beta F(r)] dr}{\int_0^{r_0} Exp[\beta F(r)] dr}=\frac{Exp[-\beta r_0^2]-1}{\sqrt{\beta \pi}\,erf(\sqrt{r_0\beta})}
\end{equation}
where $\left<\sigma_{IS}\right > \rightarrow r_0/2 = \sigma_{IS,max}$ at high temperature.

In Figure \ref{fig:Caging}B we plot $f_{caged}$ vs $\left<\sigma_{IS}\right >/ \sigma_{IS,max}$ for the above model. We see that $f_{caged} \approx 0.85$ when $\left<\sigma_{IS}\right >/ \sigma_{IS,max}$ is just greater than $0.8$, as it is at $T_A$.

\newpage

\subsection*{Characteristic Temperatures}
\begin{table}[hbt!]
	\caption {Characteristic Temperatures}
	
	\begin{tabular}{l|l|l|l|l}
		
		
		\hline\\
		\textbf{Molecule}                               & $T_A$ {[}K{]} & $T_B$ {[}K{]}  & $T_A$? {[}K{]} & $T_B$? {[}K{]}\\
		\hline
		\hline
		propylene carbonate                    & 291$^a$ & 209$^a$ & &        \\
		& 294 \cite{Stickel:1995du}        & 200 \cite{Stickel:1995du}        & &\\
		&           & 187  \cite{Domschke:2011jf,Schonhals:1993ei,Du:1994kx}      & &\\
		&            & 182 \cite{Wuttke.2000}& &           \\
		&             & 196 \cite{Borjesson:1990} & &\\
		&             & 210$^b$ & &\\
		\hline 
		average      & 292$\pm$2&   195$\pm$10    &            &            \\
		\hline
		\hline
		
		propylene glycol                       & 321  \cite{Elmatad:2009ec}      &  & &280 \cite{Stickel:Thesis} \\
		& 292 \cite{Schonhals:1993ei}           &  & &239 \cite{Lunkenheimer:2012iz}\\
		&            & & &251\cite{Schonhals:1993ei}\\
		&            & & &248$^b$\\
		\hline		
		average                                & 305$\pm$20          &            & &255$\pm$18\\
		\hline
		\hline
		
		glycerol &  & 310 \cite{Rossler:1994}&338 \cite{Elmatad:2009ec}&\\
		&& 288\cite{Adichtchev:2003} &330\cite{Rzoska:2019} &\\
		&& 285\cite{Stickel:Thesis}     & 288\cite{Schonhals:1993ei}* &\\
		&&295\cite{Rzoska:2019}&&\\
		&&283\cite{Schroter:2000fq}&&\\
		\hline
		average      & &294$\pm$13 &    334$\pm$6  &            \\
		\hline
		\hline
		
		ortho terphenyl                        & 455 \cite{Hansen:1997ic}       & 290\cite{Hansen:1997ic,Rossler:1994,Steffen:1994,Stickel:Thesis}&341\cite{Elmatad:2009ec}  &      \\
		&  & 285 \cite{Gottke:2001jv}        &357\cite{Elmatad:2009ec}  &\\
		& & 293 \cite{Petry:1991}           &340 \cite{Elmatad:2009ec} &\\
		&&291\cite{Garcia:2012}&&\\
		&&279$^b$&&\\
		\hline
		average      & &288$\pm$5      &346$\pm$10 &            \\
		\hline
		\hline
		sorbitol  & & 340\cite{Rzoska:2019}& &\\
		&            &351$^b$ & &\\
		\hline
		average      &&345$\pm$8&&\\
		\hline
		\hline
		Kob-Andersen  & 0.84 $^c$& 0.435\cite{Kob.1995bv}& &\\
		&            & & &\\
		\hline
		average      &0.84$\pm$.09&0.435&&\\
		\hline
		\hline
		\label{tab:charT}
	\end{tabular}
	\\
	\footnotesize{ $a$ $T_A$ and $T_B$ obtained from data of Leukenheimer et al. \cite{Lunkenheimer:2012iz}.\\ $b$ $T_B$ values estimated as the temperature where $\alpha$ and $\beta$ relaxation processes merge.\\ * This temperature was designated at $T_A$ in ref \cite{Schonhals:1993ei} but corresponds to $T_B$ as determined by other authors, so it is not included in the average for $T_A$.\\$c$ as determined from Fig.\ref{fig:KAtau}}\\
\end{table}

Table \ref{tab:charT} displays characteristic temperatures $T_A$ and $T_B$ obtained from literature sources. Where $T_B$ values have been determined, it is generally found to be statistically indistinguishable from the mode-coupling critical temperature $T_c$, although the two are found by distinct methods. Here we take them to be functionally interchangeable and include $T_c$ values in the average and standard deviation we find for $T_B$. For sorbitol, where little corroborating data for $T_A$ and $T_B$ values can be found, but high quality $\tau_{\beta, JG}$ data is available well above $T_g$ \cite{Fujima:2002fr}, we estimated $T_B$ as the temperature where $\tau_{\alpha}$ and $\tau_{\beta, JG}$ bifurcate. A similar analysis was carried out for PC and OTP, and these  values are marked with a superscript $b$ in Table \ref{tab:charT}. In cases marked with a superscript $a$, characteristic temperatures were determined from derivative analysis of published relaxation data for PC \cite{Ngai:2001di}.

In several cases, the temperature at which $P_0=0.24$ appears to have been misidentified as $T_A$ or $T_B$. These are labeled as ``$T_A$?'' or ``$T_B$?''. 

\newpage

\subsection*{Coordination Number}

\begin{table}[hbt!]
	\caption {Coordination Numbers}
\begin{tabular}{l|l|l}
\hline\\
\textbf{Molecule}& T [K] & z\\	\hline	\hline
		propylene carbonate$^a$   & 200&12.46\\\hline
		   & 220&12.36\\\hline
		   & 240&12.44\\\hline
		   & 260&12.35\\\hline
		 & 280&12.46\\\hline
		 & 300&12.43\\\hline
		 & \textbf{Value Used} &12.4\\\hline
		 
		propylene glycol \cite{Ferreira.2017} &298 &13\\ \hline
		 & \textbf{Value Used} &13\\\hline
		glycerol\cite{Cicerone:2015wn} &300&13.09\\\hline
		&350&13.59\\ \hline
		&400&12.98\\\hline
		&450&13.17\\\hline
		 & \textbf{Value Used} &13.3\\\hline
		
		ortho terphenyl (LW)$^a$& 275 &10.86\\ \hline
		& 300 &10.75\\ \hline
		& 400 &10.5\\ \hline
		&  500&10.21\\ \hline
		ortho terphenyl (atomistic)\cite{Hung:2019fm}& 267 &13.3\\ \hline
		 & \textbf{Value Used} &11.5\\\hline

		sorbitol$^a$  & 373&12.9 \\ \hline
		 & \textbf{Value Used} &12.9\\\hline
		Kob-Andersen$^a$  & 1.0&11.25 \\ \hline
		 & \textbf{Value Used} &11.25\\\hline
	
		\hline
		\label{tab:zs}
	\end{tabular}
	\\
\footnotesize{$a$ This work.\\ }
\end{table}

Coordination numbers (z) used to construct Fig. 3 in the main text are given in Table \ref{tab:zs} and were calculated from simulated center-of-mass radial distribution functions ($g(r)$), 

\begin{equation}
	z(r_{min})=4\pi\rho \int_{0}^{r_{min}}g(r)r^2dr
\end{equation}
where $\rho$ is the average density, and $r_{min}$ is the first minimum in $g(r)$.

\begin{figure}[!h]
	\includegraphics[width=12 cm]{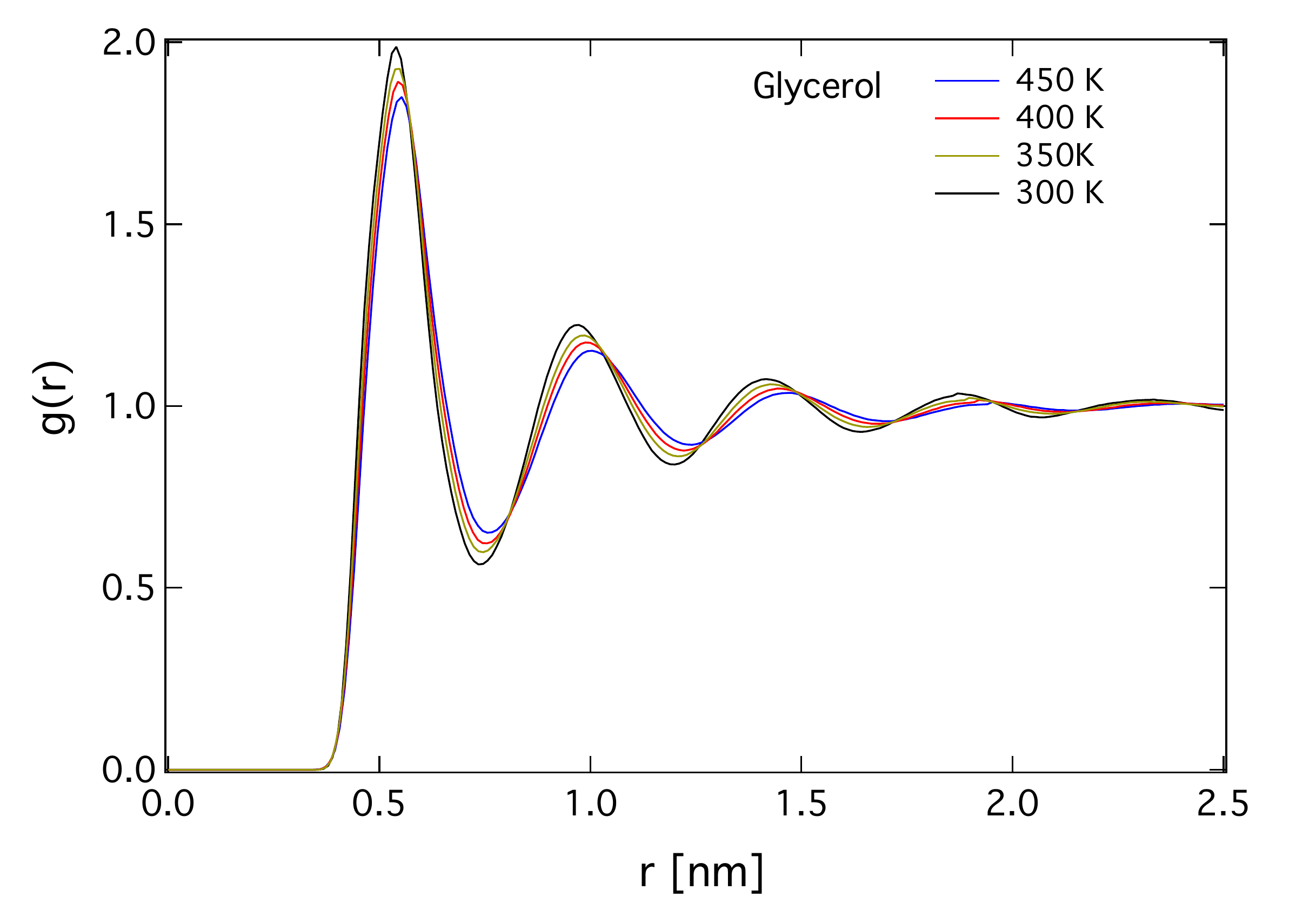}
	\caption{\footnotesize{Radial distribution functions for glycerol calculated from trajectories obtained as described in \cite{Cicerone:2015wn}.}}\label{fig:grgly}
\end{figure}

For glycerol, simulations were performed as described in ref. \cite{Cicerone:2015wn}, and $g(r)$ was calculated at 50 K intervals between 300 K and 450 K as shown in Fig. \ref{fig:grgly}. 

PC and sorbitol simulation consisted of 400 and 200 molecules, respectively, initially constructed using Packmol \cite{Martinez2003PackingDocking}. NPT simulations at 1 bar were performed at each temperature using a Monte Carlo Barostat and were run for 10 ns, and subsequent NVT simulations were run for 100 ns. 
All simulations were run using a 1 fs timestep using the GPU-accelerated OpenMM software.\cite{Eastman2017OpenMMDynamicsb}. The \textit{ab initio}, polarizable SAPT-FF force field was used for propylene carbonate, while the CHARMM36 force field was used for sorbitol \cite{McDaniel2013Physically-motivatedTheory, Huang2013CHARMM36Data}. The procedure for constructing the force field for propylene carbonate was described elsewhere \cite{Stoppelman.2022}. A dual-Langevin thermostat was used for treating the Drude oscillator degrees of freedom for the propylene carbonate simulations, with a 1 ps$^{-1}$ friction coefficient \cite{Lamoureux2003ModelingAlgorithm}. A standard Langevin thermostat was used for sorbitol, also with a 1 ps$^{-1}$ friction coefficient. The particle-mesh Ewald method (PME) was used for modeling electrostatics in all simulations, and a 1.4 nm cutoff is used for handling van der Waals (VDWs) interactions \cite{Essmann1995AMethod}.

OTP was simulated using the Lewis-Wahnstr{\"o}m (LW) 3-site model.\cite{Lewis1994Molecular-dynamicsiorthoi-terphenyl}, and MD simulations of 200 OTP molecules were run using the GROMACS software package \cite{Abraham2015Gromacs:Supercomputers}. 
The LINCS algorithm was used to keep bonds and angles constrained \cite{Hess2007P-LINCSSimulation}.
NPT simulations were run at 1 bar for 5 ns using the Parrinello-Rahman barostat \cite{Parrinello1998PolymorphicMethod}, and subsequent NVT simulations were then performed for 5 ns using a Nos{\'e}-Hoover thermostat with a time constant of 1 ps.\cite{Hoover1985CanonicalDistributions}
These LW OTP simulations yielded center-of-mass z values of  16.3, 15.6, and 15.2 at 300 K, 400 K, and 500 K, respectively, in agreement with LW model calculations from Boue et al. \cite{Boue:2011}. 

Using z in the range (15 to 16) results in $P_0$, $P_1$, and $P_2$, values far from 0.24. Rather, we find agreement between the behavior of OTP and the other molecules if we use a z value for OTP phenyl rings rather than molecular centers of mass, and this is what we have reported in Table \ref{tab:zs}. We use a phenyl ring z value of 11.5, which is intermediate between values obtained in LW OTP and atomistic OTP simulations performed by Hung et al. \cite{Hung:2019fm}. 

\begin{figure}[!h]
		\includegraphics[width=12 cm]{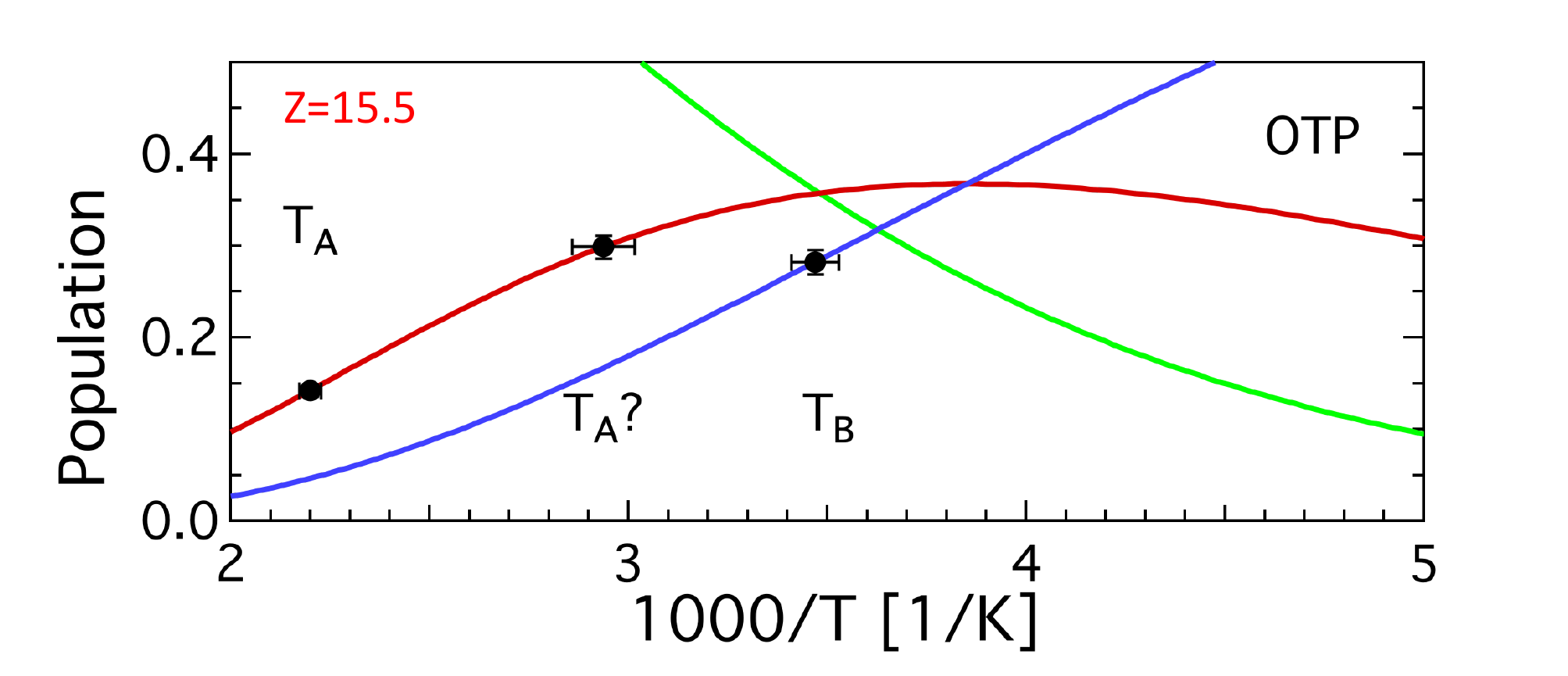}
	\caption{\footnotesize{Estimated populations of environments with 0, 1, and $\geq 2$ excitations in the first shell as a function of temperature for OTP, using the molecular center of mass coordination number, z = 15.5.}\label{fig:OTPz15}}
\end{figure}

Figure \ref{fig:OTPz15} shows that characteristic temperatures do not fall at iso-population points when using the OTP molecular center of mass coordination number, z=15.5 to calculate $P_0$, $P_1$, and $P_2$ as they do when we use the phenyl ring coordination number, as in Fig. 3B of the main text. This suggests that it matters only whether a phenyl group in the first shell of another phenyl group is involved in an excitation. We can understand this if we assume that excitations involve only a small number of atoms. For example, if two OTP molecules are positioned so that their central rings are adjacent, an excitation involving a few atoms in the central ring of molecule 1 is likely to impact the dynamic environment of the central ring  of molecule 2, whereas an excitation on a distal ring of molecule 1 would be unlikely to do so. This sub-molecular hypothesis is consistent with what is known of these phenomena from LJ simulations and with comparisons of H-atom and COM intermediate scattering functions \cite{Stoppelman.2022}. In LJ simulations, only a few particles are simultaneously  involved in an excitation, and LJ particles are more representative of atoms than of entire molecules. Also, analysis of H-atom and center-of-mass intermediate scattering functions in glycerol \cite{Cicerone:2015wn} show that COM motion qualifying as hops is relatively rare compared to hop prevalence from incoherent H-atom motion.

\newpage

\subsection*{Propylene Glycol and Sorbitol Dynamic Environment Populations}
\begin{figure}[h!]
	\begin{center}
		\includegraphics[width=12 cm]{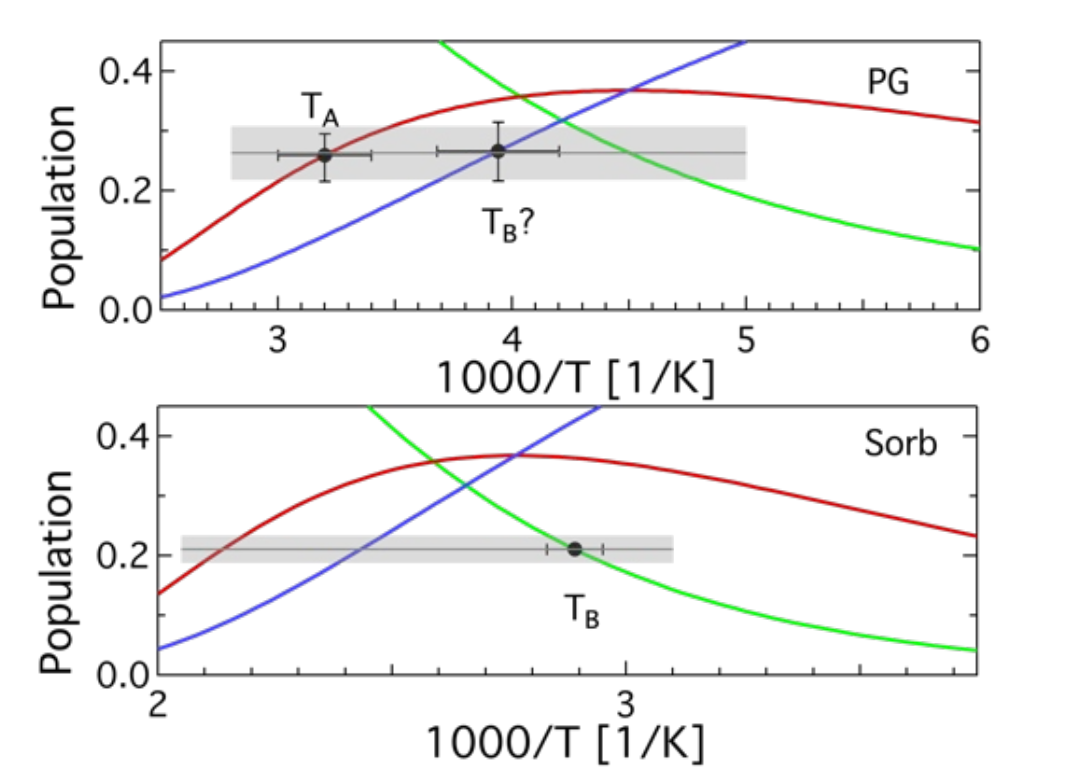}
		\caption{Estimated populations of environments with 0, 1, and $\geq 2$ excitations in the first shell as a function of temperature for liquids indicated. Green, red, and blue lines indicate $P_2$, $P_1$, and $P_0$ populations, respectively. $T_A$ and $T_B$ values, where established, are marked by solid black circles. Each of these temperatures coincides with a population of $0.24\pm 0.2$ for one of the three dynamic environments, $P_2$, $P_1$, or $P_0$.}\label{fig:SIRegimes}
	\end{center}
\end{figure}

Due to length limits in the main manuscript, we present estimated dynamic populations $P_0$, $P_1$, and $P_2$ in propylene glycol (PG) and sorbitol (sorb) here in Figure \ref{fig:SIRegimes}. As in Fig. 3 of the main text, we estimate the populations of these distinct environments by assuming that excitations are randomly distributed in space and that $P_0= 1$ at $T=0$. Under these assumptions, $dP_0/d\Phi_0=-(z+1)\,P_0$, where z is the coordination number, and $dP_i/d\Phi_0=(z+1)\,(P_{i-1}-P_i)$ for $i>0$. 

\newpage

\subsection*{Kob Andersen LJ: Simulation, Identification of $T_A$  and Identification of Excitations}

\textbf{Kob-Andersen (KA) binary LJ simulations} were performed with 10,000 particles at the standard choice of 80\%/20\% A/B particles and reduced density $\rho=1.2$. In the KA mixture, the A/B particles interact with LJ pair potentials given by $\sigma_{BB}/\sigma_{AA}$=0.88, $\sigma_{AB}/\sigma_{AA}$=0.8, $\epsilon_{BB}/\epsilon_{AA}$=0.5, and $\epsilon_{AB}/\epsilon_{AA}$=1.5. KA simulations were run with the OpenMM software, utilizing a Langevin thermostat and pairwise interactions cutoff at $r_{cutoff} \approx 3 \sigma_{AA}$. Different temperature simulations were performed over the range T=0.4-1.2 in reduced temperature units. The high-temperature simulations were equilibrated and run for between 100,000-500,000 reduced time units, while the lowest-temperature simulation was run for up to 2,000,000,000 reduced time units. 

\begin{figure}[!h]
		\includegraphics[width=9 cm]{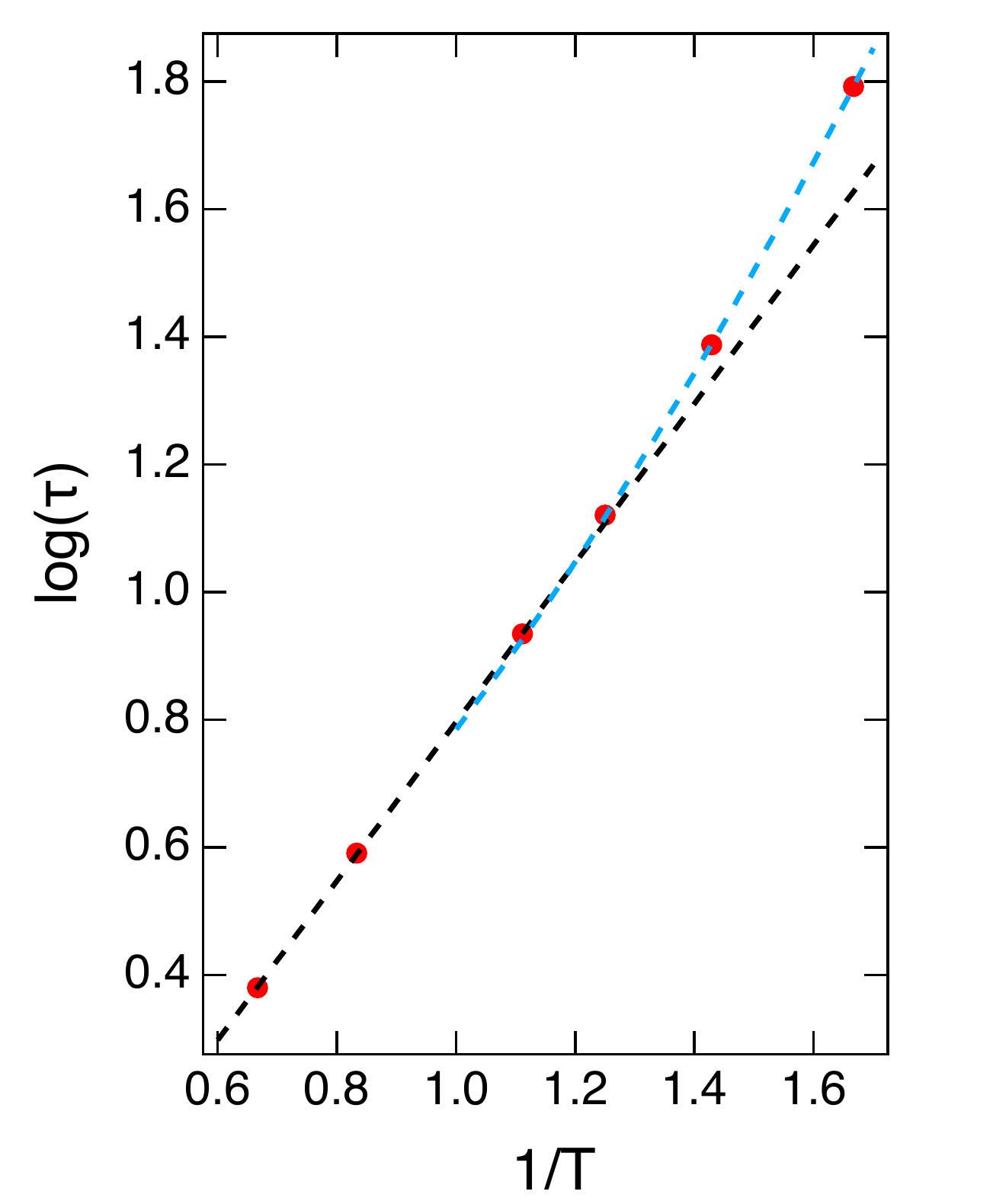}
	\caption{\footnotesize{Relaxation times for A particles in the KA system, determined through $F_s(q,t)$ The black dashed line is a linear fit and the blue dashed line is quadratic in ($1/T-1/T_A$) with $T_A=0.84\pm0.09$.}\label{fig:KAtau}}
\end{figure}

Relaxation times determined as the 1/e point of $F_s(q,t)$ at $q_{max}$ for A particles are shown in Fig. \ref{fig:KAtau}. From this data we identify $T_A=0.84\pm.0.9$. Above $T_A$ we find an Arrhenius fit but a super-Arrhenius tperature dependence for $T<T_A$. The blue dashed line at $T<T_A$ is a quadratic fit.

\textbf{Hops were identified in KA trajectories} by first characterizing the motion of each particle at each point in time based on the standard deviation, $stdev$, of the particle's position over the course of the previous 1.52 ps.


$\Phi_{0}$, determined from analysis of $F_s(q,t)$ of A particles, is used to set the fraction of particles that will be considered as being involved in an excitation. The windowed standard deviation values for each particle in each frame are arranged in increasing order, and the fraction $\Phi_0$ with the highest $stdev$ values are labeled as being in involved in an excitation. 

Particles are identified as belonging to $P_0$, $P_1$, and $P_2$ environments at each time point by examining the labels of the particles in its first solvation shell, defined by the interaction radius, $R_{AA}$ = 1.42 $\sigma$. Each particle starts with a label of 1 if involved in an excitation or 0 otherwise. The environment experienced by the center particle is determined by the sum of the excitation labels of its neighbor particles and its own label. If there are no excitations in its first shell, the environment label is the same as the excitation label (0 or 1). If, however, the sum of excitation labels over the first shell is greater than or equal to 2, the particle's environment is identified as $P_2$. 

\newpage

\subsection*{Difficulties in Using Simulation to Quantify Excitations in Real Molecular Systems}

There are presently two major difficulties in using simulations to quantify excitations in molecular systems. On issue, as we have recently discussed \cite{Stoppelman.2022}, is that motion involved in excitations includes low-frequency intra-molecular vibration and intermolecular motion. The accuracy of standard force fields seems to be insufficient to describe these motions in quantitative agreement with experiment. The other issue is related to complexities in defining excitations that arise in the presence of intra-molecular bonds. 

\begin{figure}[!h]
		\includegraphics[width=15 cm]{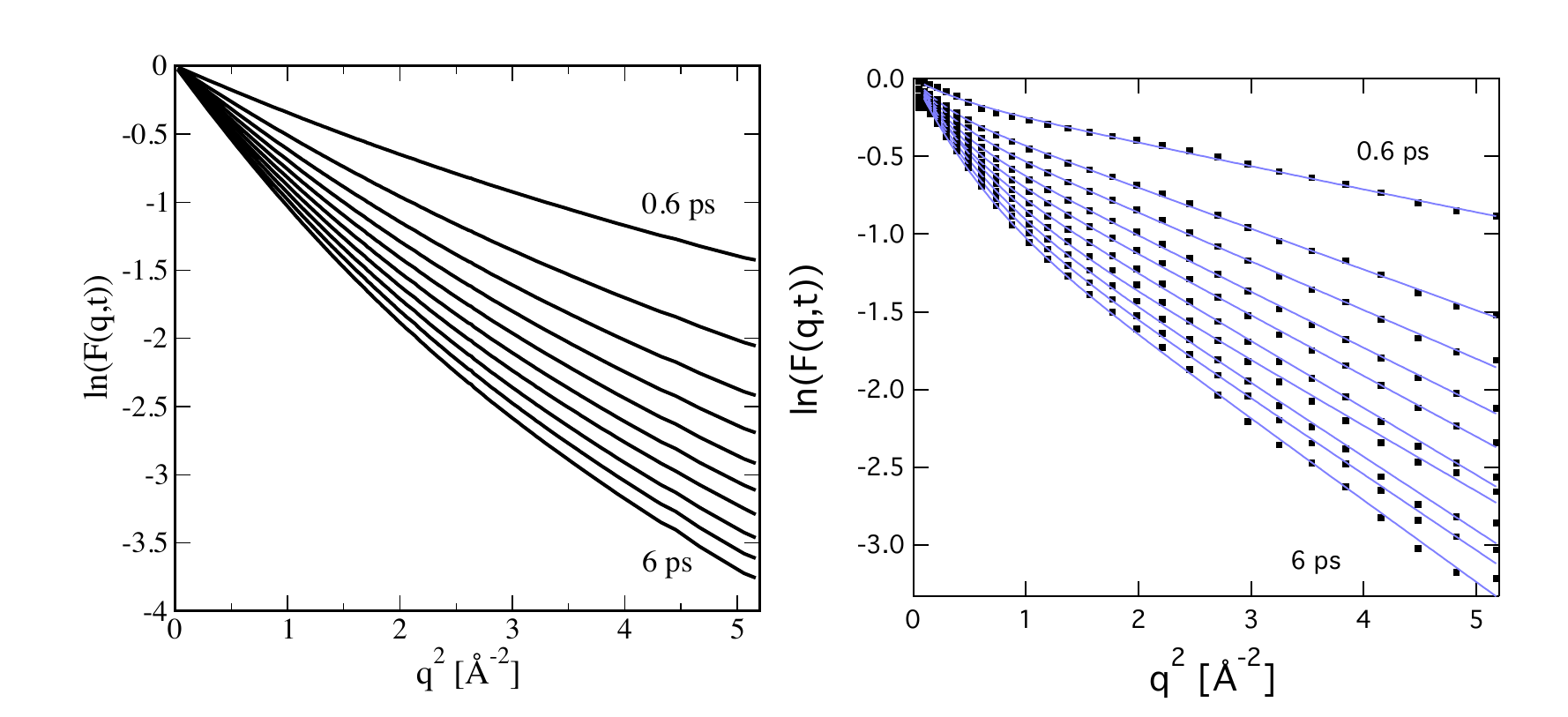}
	\caption{\footnotesize{Intermediate scattering functions for propylene carbonate at 300K. Left: $F(q,t)$ calculated from simulation trajectories as described in ref\cite{Stoppelman.2022}. Right: Points are $F(q,t)$ measured using QENS, and lines are fits to the data using equation (1) in the main manuscript.}\label{fig:Fqt_compare}}
\end{figure}

The first issue is exemplified for propylene carbonate in Fig. \ref{fig:Fqt_compare}, with data taken from ref \cite{Stoppelman.2022}. Similar results can be found in ref \cite{Cicerone:2015wn}. In both cases, the simulation predicts decay of $F_s(q,t)$ much faster than that measured experimentally at $q\geq 1\,$\AA$^{-1}$, and additional nuances of the experimental shape are not recovered. In comparing $F(q,t)$ in Fig. \ref{fig:Fqt_compare}, note that the ordinate axis extends to -4 for simulation and only to -3 for the experimental result.  Note also that the propylene carbonate force field was developed based on \textit{ab initio} calculations of intermolecular interactions, and yet still exhibited discrepancy compared to experiment.  It is thus an important research goal to improve the simulation/force field description of motions involved in excitations. We speculate that a higher accuracy treatment of intramolecular vibrations (i.e. bonded force field terms), and/or inclusion of nuclear quantum effects could improve the description of excitations. With regard to the latter, high-frequency intramolecular vibrations must certainly be treated quantum mechanically, yet excitations involve the lower frequency modes.\cite{Stoppelman.2022}  The importance of nuclear quantum effects in describing excitations thus remains an open question.

The second issue, \textit{a priori} identification of spatial extend of excitations in molecular systems is largely unexplored, but presents a challenge as touched on above in the description of finding an appropriate z value for OTP. Repeating that text: Figure \ref{fig:OTPz15} shows that characteristic temperatures do not fall at iso-population points when using the OTP molecular center of mass coordination number, z=15.5 to calculate $P_0$, $P_1$, and $P_2$ as they do when we use the phenyl ring coordination number, as in Fig. 3B of the main text. This suggests that it matters only whether a phenyl group in the first shell of another phenyl group is involved in an excitation. We can understand this if we assume that excitations involve only a small number of atoms. For example, if two OTP molecules are positioned so that their central rings are adjacent, an excitation involving a few atoms in the central ring of molecule 1 is likely to impact the dynamic environment of the central ring  of molecule 2, whereas an excitation on a distal ring of molecule 1 would be unlikely to do so. This sub-molecular hypothesis is consistent with what is known of these phenomena from LJ simulations and with comparisons of H-atom and COM intermediate scattering functions \cite{Stoppelman.2022}. In LJ simulations, only a few particles are simultaneously  involved in an excitation, and LJ particles are more representative of atoms than of entire molecules. Also, analysis of H-atom and center-of-mass intermediate scattering functions in glycerol \cite{Cicerone:2015wn} show that COM motion qualifying as hops is relatively rare compared to hop prevalence from incoherent H-atom motion. 

\newpage

\subsection*{Percolation of Mobile and Immobile Domains}

In the main text, we infer that percolation thresholds are crossed at $T_A$ and $T_B$. However, the instantaneous population of $P_1$ or $P_2$ at $T_A$ and $T_B$ are 0.24; much higher than expected for a site percolation critical fraction  ($p_c$). The amorphous systems considered here have average coordination numbers (z) ranging from 11 to 13. Crystalline f.c.c and hcp latices have coordination number (z) = 12, and $p_c$ of 0.195 and 0.199. Random packed systems are expected to have $p_c$ values similar to or slightly smaller than regular lattices with the same z. As we show below, we did identify percolation thresholds at the characteristic temperatures, but only for long-lived environments.

To test our percolation hypothesis in the KA system we directly calculated percolation for dynamic environments. We used a recursive algorithm by \citet{Edvinsson1999ClusterAlgorithm} to compute the percolation probability for a given trajectory. The code was applied for each separate environment ($P_0$, $P_1$ or $P_2$) at each time frame in simulation trajectories to determine whether there that environment forms a percolating cluster. 

To ensure we had implemented the algorithm correctly, we randomly assigned excitation labels to varying fractions of A particles in trajectories and found that each of the environments percolated at $p_c=0.2$, precisely as expected. On the other hand, when excitations were identified based on particle mobility as described above, we found that domains $P_0$, $P_1$, or $P_2$ percolate with $p_c=0.14$, suggesting that excitations are not randomly distributed but form local clusters.

\begin{figure}[!h]
	\includegraphics[width=14cm]{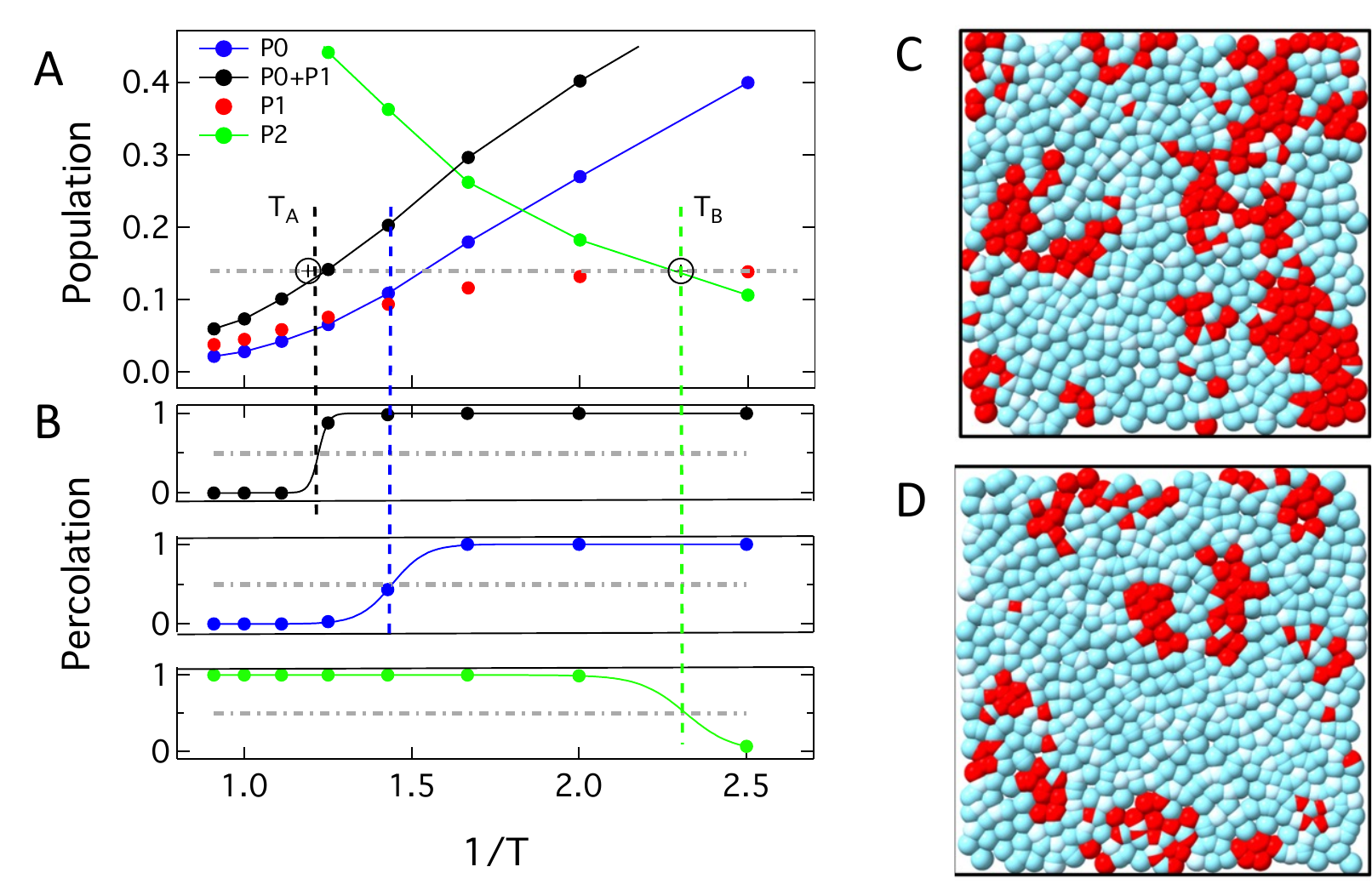}
	\caption{\footnotesize{A: Populations in environments that persist for at least 1 ps after being first defined. The open circles with plus signs indicate $T_A$ and $T_B$. The observed percolation threshold $p_c=0.14$ is indicated as a gray dashed line. B: Percolation probabilities for environments indicated in the figure legend. The solid lines are sigmoidal fits and are intended as guides to the eye. C and D: Percolating and non-percolating environments. The red particles belong to $P_2$ environments, while the cyan particles are a collection of those in $P_0$ and $P_1$. C: Example percolating environment for $P_2$, taken from a $z$-axis slice of a T = 0.5 trajectory. D:Example of a nonpercolating environment for $P_2$, taken from a $z$-axis slice of a T = 0.4 trajectory.}\label{fig:KA_perc}}
\end{figure}

We found also, that if we limit our attention to environments that persist for an additional 1 ps after being first defined, we obtain environments that percolate at the expected temperatures. $P_i$ values for environments that persist for 1 ps after being first defined are shown in Fig.\ref{fig:KA_perc}A. $p_c$ is exceeded at $T_A$ for persistently immobile particles, those that remain in $P_0$ or $P_1$ domains. Likewise, the population of mobile particles, those that persist in $P_2$ environments, drops below $p_c$ at $T_B$.

The three panels in Fig. \ref{fig:KA_perc}B show percolation probabilities for environments. The top of these three panels shows percolation probabilities of all ``immobile'' environments comprising particles in either $P_0$ or $P_1$ environments. It is immobile this, rather than simply $P_1$ that percolates at $T_A$. The middle of these three bottom panels shows that a percolation of solid-like environments occurs at a temperature intermediate between $T_A$ and $T_B$. The bottom panel shows that mobile $P_2$ domains cease to percolate at $T_B$. 

We show in Figure \ref{fig:KA_perc}C and D, examples of percolating and nonpercolating environment for $P_2$ at T = 0.5 and T=0.4 respectively. The images are 2D slices along the $z$-axis of $\sim 1.3\sigma$ in depth from our 3D trajectories. All particles from the slice were then projected onto a single plane for visualization purposes. 


\newpage

\subsection*{Using KA Percolation Results as Surrogate for Molecular Systems}

We have previously shown that, when normalized by the hydrodynamic radius, the molecular motion of propylene carbonate (PC) associated with inherent state and interbasin hop motion scales precisely with that of the KA model \cite{Cicerone:2017eo}. We understand this as an indication that the essential physics that underlies this motion is identical for both systems when the molecular system is scaled appropriately to mostly average over intramolecular motion. The other liquids in this work, PG, glyc, sorb, and OTP, exhibit the same dynamic hallmarks as KA when the former are scaled by the hydrodynamic radius \cite{Cicerone:2014dk}. As described next, we have also demonstrated dynamic mapping in more recent work.

KA, and other simple liquid models exhibit density scaling of dynamic and thermodynamic properties in that they are invariant at constant values of $\rho^{\gamma_{DS}}/T$ where $\gamma_{DS} = \gamma_G$, and $\gamma_G$ is the Gr$\ddot{\textrm{u}}$neisen parameter. We have shown that the dynamics of PC also conform to this law with $\gamma_{DS} = \gamma_G$ only when we average appropriately over intramolecular motion. Specifically when we average over motion faster than IB transitions. Thus, when intramolecular dynamics are spatially and temporally averaged so that the focus is on IB transitions, KA makes an excellent model for PC. We have yet to demonstrate this in the other liquids analyzed here, but have every reason to believe we will find identical results, given the IB and IS motion lengthscale correspondence we have shown previously \cite{Cicerone:2014dk,Cicerone:2017eo}.  

The probability that each of molecular liquids would randomly exhibit the same $P_i=0.24$ at $T_A$ and $T_B$ as the KA system is on the order of $10^{-7}$. Given this, and the dynamic mapping between KA and molecular liquids discussed above, we are comfortable using KA as a model system for these liquids. 

Physically, the site percolation threshold, $p_c$ depends on the coordination number ($z$) and fractal dimension ($\xi$) of site clusters. The $z$ values for each of these liquids and KA are similar. While $\xi$ for each of these systems is unknown, we can infer their similarity by the similar behavior of each system in Fig. 3A-D of the main text and Fig. \ref{fig:SIRegimes}.


\newpage



\newpage

%